\documentclass{article}
\usepackage{spconf,amsmath,graphicx}

\def\L{{\cal L}}

\usepackage{cite}
\newcommand{\blskip}{\baselineskip 10.15pt}
\usepackage{bm}
\def\T{{\top}}
\def\A{{\bm{A}}}
\def\E{{\bm{E}}}
\def\X{{\bm{X}}}
\def\P{{\bm{P}}}
\def\a{{\bm{a}}}
\def\e{{\bm{e}}}
\def\x{{\bm{x}}}
\def\h{{\bm{h}}}
\def\z{{\bm{z}}}
\def\p{{\bm{p}}}
\def\orah{\overrightarrow{\mathstrut\bm{h}}}
\def\olah{\overleftarrow{\mathstrut\bm{h}}}
\def\embed{{\mathtt{embed}}}
\def\concat{{\mathtt{concat}}}
\def\fwlstm{{\mathtt{fwlstm}}}
\def\bwlstm{{\mathtt{bwlstm}}}
\def\linear{{\mathtt{linear}}}
\def\softmax{{\mathtt{softmax}}}
\def\CTC{{\mathrm{CTC}}}
\def\Att{{\mathrm{Att}}}
\def\LM{{\mathrm{LM}}}
\def\CB{{\mathrm{CB}}}

\title{BLSTM-BASED CONFIDENCE ESTIMATION FOR END-TO-END SPEECH RECOGNITION}
\name{Atsunori Ogawa, Naohiro Tawara, Takatomo Kano, and Marc Delcroix}
\address{NTT Corporation, Japan}
\begin{document}
\ninept
\blskip
\maketitle
\begin{abstract}
Confidence estimation,
in which we
estimate the reliability of each recognized token
(e.g., word, sub-word, and character)
in automatic speech recognition (ASR) hypotheses
and detect incorrectly recognized tokens,
is an important function for developing ASR applications.
In this study,
we perform confidence estimation
for end-to-end (E2E) ASR hypotheses.
Recent E2E ASR systems show high performance
(e.g., around 5\% token error rates)
for
various
ASR
tasks.
%
In such situations,
confidence estimation becomes difficult
since we need to detect
infrequent incorrect tokens
from
mostly
correct token sequences.
To tackle this 
imbalanced dataset
problem,
we employ a bidirectional long short-term memory
(BLSTM)-based model as
a
strong
binary-class
(correct/incorrect)
sequence labeler
that is
trained with a
class balancing objective.
We experimentally confirmed that,
by
utilizing
several types of ASR decoding scores
as
its
auxiliary features,
the model steadily shows
high confidence estimation performance
under
highly imbalanced
settings.
We also confirmed that
the BLSTM-based model outperforms
Transformer-based
confidence estimation
models,
which
greatly
underestimate
incorrect tokens.
\end{abstract}
%
\section{Introduction}
\label{sec_intro}
%
Based on the recent introduction of
state-of-the-art
neural network (NN) modeling,
the performance of automatic speech recognition (ASR)
has been greatly improved
\cite{Hinton_IEEESPM2012,Yu_Springer2015}.
However,
since
errors are essentially unavoidable in ASR,
confidence estimation
\cite{Jiang_SPECOM2005},
which is
a
function
that estimates the reliability of each recognized token
(e.g., word, sub-word, and character)
in ASR hypotheses,
is important for developing such ASR applications
as
spoken dialogue systems
\cite{San-Segundo_ICASSP2001}
and smart speakers
\cite{Swarup_IS2019}.

Many confidence estimation methods have been proposed
and they can be roughly
categorized into three types of approaches
\cite{Jiang_SPECOM2005}.
The first
is the utterance verification approach,
which formulates confidence estimation
as a statistical hypothesis testing problem
\cite{Lleida_IEEETSAP2000}.
The second
category
approach
is based on token posterior probabilities
calculated over ASR hypotheses
in the form of
$N$-best lists,
lattices,
and confusion networks
\cite{San-Segundo_ICASSP2001,Wessel_IEEETASLP2001,Mangu_CSL2000}.
This approach is easy-to-use
since it works on ASR hypotheses
without
any extra models.
The third category consists of
classifier-based approaches
\cite{San-Segundo_ICASSP2001,Fayolle_IS2010,Seigel_IS2011,Yu_IEEETASLP2011,Schaaf_ICASSP1997,Weintraub_ICASSP1997,Tam_ICASSP2014,Kalgaonkar_ICASSP2015,Ogawa_ICASSP2015,Ogawa_SPECOM2017,Del-Agua_IEEEACMTASLP2018}.
Since this last category provides superior performance,
we focus on such approaches in this study.

Confidence estimation is a binary-class
(correct/incorrect)
sequence labeling problem
for which various types of sequence labeling models have been applied,
for example,
conditional random fields (CRFs)
\cite{Fayolle_IS2010,Seigel_IS2011,Yu_IEEETASLP2011,Ogawa_ICASSP2015,Ogawa_SPECOM2017},
multilayer perceptrons (MLPs)
\cite{San-Segundo_ICASSP2001,Fayolle_IS2010,Seigel_IS2011,Yu_IEEETASLP2011,Schaaf_ICASSP1997,Weintraub_ICASSP1997,Tam_ICASSP2014,Kalgaonkar_ICASSP2015,Ogawa_SPECOM2017},
and recurrent NNs (RNNs)
\cite{Kalgaonkar_ICASSP2015,Ogawa_ICASSP2015,Ogawa_SPECOM2017,Del-Agua_IEEEACMTASLP2018}.
Among these models,
the bidirectional long short-term memory (BLSTM)
\cite{Hochreiter_NeuralComput1997,Schuster_IEEETSP1997}
based model
shows especially high
performance
\cite{Ogawa_ICASSP2015,Ogawa_SPECOM2017,Del-Agua_IEEEACMTASLP2018}
since it can utilize
forward and backward
longer input contextual features
than the other models
to estimate the confidence score of each token
in a token sequence
\cite{Ogawa_ICASSP2015,Ogawa_SPECOM2017}.
We input
a recognized token sequence
(ASR hypothesis)
with
a corresponding
auxiliary feature vector sequence
to these models
and
obtain
a binary-class
(correct/incorrect, i.e., two-dimensional)
posterior probability
vector sequence for the input token sequence.
Then
by applying thresholding
to the obtained
probability vector sequence,
we can detect incorrectly recognized tokens from the input token sequence.
Various types of auxiliary features
can be
extracted
with the ASR decoding process
and post-processing.

Recently,
end-to-end (E2E) ASR systems
\cite{Chorowski_arXiv2014,Chan_ICASSP2016,Watanabe_IS2018,Ott_NAACLHLT2019}
have shown
very high ASR performance that is close to
(or higher than)
that of
deep NN (DNN) / hidden Markov model (HMM) hybrid ASR systems
\cite{Karita_ASRU2019}.
However,
the confidence estimation methods described above
have only been
applied to Gaussian mixture model (GMM)/HMM or DNN/HMM hybrid ASR systems.
To the best of our knowledge,
only
a few studies
have recently been
conducted on confidence estimation for E2E ASR systems
\cite{Woodward_IS2020,Kumar_IS2020}.

In this study,
as with \cite{Woodward_IS2020,Kumar_IS2020},
we perform confidence estimation for an E2E ASR system
\cite{Watanabe_IS2018}.
As described above,
recent E2E ASR systems show high performance
(e.g., around 5\% token error rates)
for
various
ASR benchmark tasks
\cite{Karita_ASRU2019}.
In this situation,
confidence estimation becomes difficult
since we need to detect infrequent incorrect tokens
from mostly correct token sequences.
To tackle this 
imbalanced dataset
problem,
we employ a 
BLSTM-based model as
a
strong
binary-class
(correct/incorrect)
sequence labeler
(Section~\ref{ssec_blstm})
together with efficient auxiliary features
available with the ASR decoding process
(Section~\ref{ssec_aux}).
We also use class-balanced loss
\cite{Cui_CVPR2019}
for stable model training with an imbalanced dataset
(Section~\ref{ssec_cbloss})
and a Transformer-based sequence labeler
\cite{Vaswani_NIPS2017,Guo_NAACLHLT2019,Yan_arXiv2019}
as a more advanced confidence estimation model
(Section~\ref{ssec_trans}).
We conduct experiments under highly imbalanced
settings
(Section~\ref{sec_exp}).
We believe that,
along with \cite{Woodward_IS2020,Kumar_IS2020},
our experimental results
will be very informative in the research area of confidence estimation.
%
\section{Confidence estimation method}
\label{sec_ce}
%
We describe
a BLSTM-based confidence estimation model,
the auxiliary features used in it,
a method that stabilizes the model training,
and a Transformer-based model.
%
\subsection{BLSTM-based model}
\label{ssec_blstm}
%
Let
$W$ $=$ $w_1,\cdots,w_T$
be
an ASR
hypothesis (token sequence)
of length (number of tokens) $T$
provided by
an E2E ASR system.
$w_t$ is a token recognized at time step $t$.
A token can be a word, a sub-word, or a character.
$\A$ $=$ $\a_1,\cdots,\a_T$
is an auxiliary feature vector sequence
that
corresponds
to $W$.
$\a_t$ is
the auxiliary feature vector 
attached to
$w_t$
and
consists of various types of scores obtained with the ASR decoding process
(detailed in Section~\ref{ssec_aux}).
$y_t$ is the binary-class (0 or 1) symbol.
$y_t$ $=$ $0$ denotes that $w_t$ is a correctly recognized token,
and
$y_t$ $=$ $1$ denotes that $w_t$ is an incorrectly recognized token.
Given $W$ and
$\A$,
a confidence estimation model
probabilistically
estimates whether
$y_t$ is $0$ or $1$
for
every time step.

\begin{figure}[t!]
  \centering
  \includegraphics[width=0.825\linewidth]{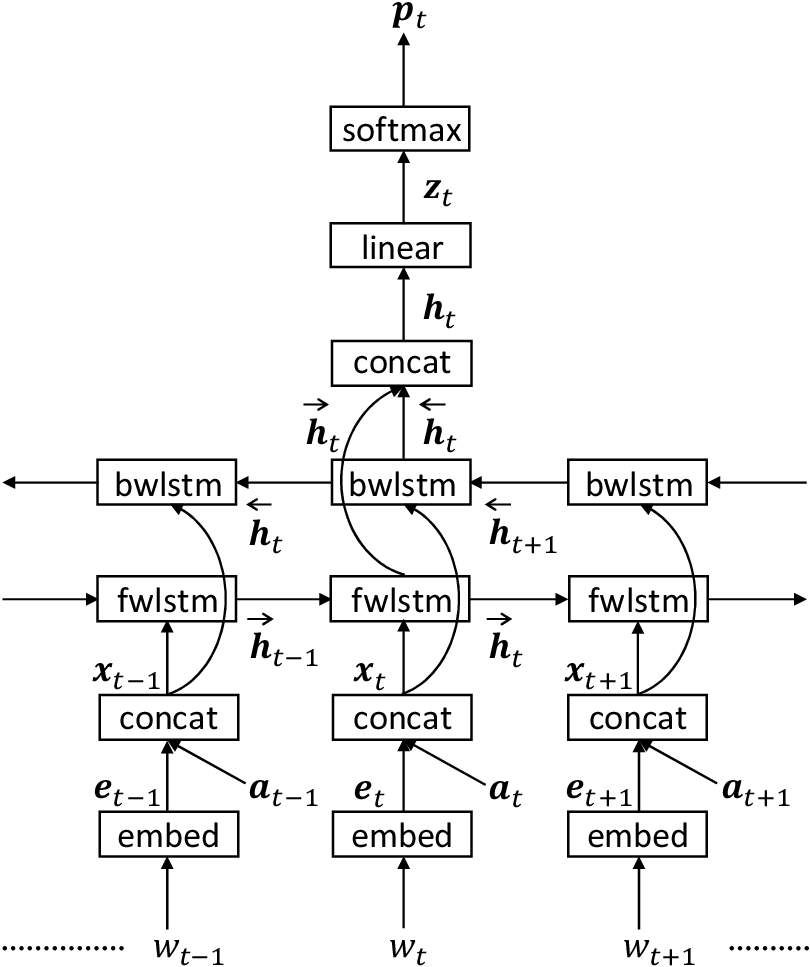}
  \caption{\blskip BLSTM-based confidence estimation model.}
  \label{fig_blstm}
  \vspace{0.2mm}
\end{figure}

Figure~\ref{fig_blstm}
shows a BLSTM-based model that performs
the above confidence estimation
(binary-class sequence labeling)
procedure
(in this figure and in the following explanation,
we assume that the number of LSTM layers in the model is one
for simplicity).
Each token $w_t$ in the given $W$ is converted to 
token embedding $\e_t$ $=$ $\embed(w_t)$,
and token embedding sequence $\E$ $=$ $\e_1,\cdots,\e_T$
is obtained.
$\E$ is concatenated with $\A$ for all the time steps,
and feature vector sequence
$\X$ $=$ $\x_1,\cdots,\x_T$
is obtained,
where $\x_t$ $=$ $\concat(\e_t,\a_t)$.
Given a forward (backward)
partial
feature vector
sequence
$\x_{1:t}$ $=$ $\x_1,\cdots,\x_t$
($\x_{T:t}$ $=$ $\x_T,\cdots,\x_t$),
a forward (backward) LSTM unit
recursively estimates forward (backward) hidden state vectors
and
provides a hidden state vector at time step $t$ as,
\begin{align}
\orah_t &= \fwlstm(\x_t,\orah_{t-1}), \\
\olah_t &= \bwlstm(\x_t,\olah_{t+1}).
\end{align}
These hidden state vectors
are concatenated as,
\begin{equation}
\h_{t} = \concat(\orah_t,\olah_t).
\end{equation}
$\h_t$ is then input into a linear layer,
followed by the softmax activation function,
and
finally,
we obtain a
binary-class
(correct/incorrect, i.e., two-dimensional)
posterior probability vector
for $w_t$ as,
\begin{align}
\z_t &= \linear(\h_t), \\
\p_t &= \softmax(\z_t),
\end{align}
where,
\begin{equation}
\p_t=[p_{0,t},p_{1,t}]^{\T}
=[P(y_t\!=\!0\,|\,W,\A),P(y_t\!=\!1\,|\,W,\A)]^{\T},
\label{eq_pt}
\end{equation}
and $\sum_cP(y_t=c\,|\,W,\A)=1$.
By repeating this procedure for all the time steps,
the model provides a posterior probability sequence
$\P=\p_1,\cdots,\p_T$ for $W$.
Then
by applying thresholding to $\P$,
we can detect the incorrectly recognized tokens from $W$.
In the model training,
we use
pairs of feature vector sequences
and corresponding binary-class (correct/incorrect) symbol sequences
(teacher label sequences) as the pair data.
%
%
\subsection{Auxiliary features}
\label{ssec_aux}
%
As the E2E ASR system,
in this study,
we employ ESPnet
\cite{Watanabe_IS2018},
which is a widely-used E2E speech processing
toolkit.
It
supports a
state-of-the-art
Transformer-based attention encoder-decoder model
\cite{Vaswani_NIPS2017}
and shows high ASR performance
for various benchmark ASR tasks
\cite{Karita_ASRU2019}.
One key feature of ESPnet is that
it employs
a connectionist temporal classification (CTC) module
\cite{Graves_ICML2014}
as a sub-decoder
along with an attention-based decoder
on top of an attention-based encoder
\cite{Kim_ICASSP2017},
that is,
ESPnet is a hybrid CTC/attention-based system
\cite{Watanabe_IEEEACMTASLP2017,Watanabe_IS2018}.
In
multi-objective style
model training,
CTC loss helps to generate monotonic attentions
between input acoustic feature vector sequences
and
corresponding
output token sequences
and stabilizes/accelerates the training
\cite{Kim_ICASSP2017,Watanabe_IEEEACMTASLP2017}.

In the decoding (beam search),
CTC and attention-based scores are jointly used.
In addition,
LSTM
language model (LSTMLM) scores are used
with shallow fusion fashion
\cite{Hori_IS2017}.
The
decoding
score for a partial hypothesis (token sequence)
$w_{1:t}$ $=$ $w_1,\cdots,w_t$
is obtained as,
%
\begin{equation}
a(w_{1:t})
= \lambda\,a_{\CTC}(w_{1:t})
+ (1-\lambda)\,a_{\Att}(w_{1:t})
+ \gamma\,a_{\LM}(w_{1:t}),
\label{eq_wsum}
\end{equation}
%
where
$a_{\CTC}(w_{1:t})$,
$a_{\Att}(w_{1:t})$,
and
$a_{\LM}(w_{1:t})$
are
respectively
the CTC, attention, and LSTMLM scores
for $w_{1:t}$,
and $\lambda$ ($0$ $\leq$ $\lambda$ $\leq$ $1$)
and $\gamma$ ($\gamma$ $\geq$ $0$)
are the tunable score weighting parameters.
Weighted sum score $a(w_{1:t})$ is used
for
the decoding.
In this study,
we use these four scores for token $w_t$
as the auxiliary features, i.e.,
\begin{equation}
\a_t=[\,a_{\CTC}(w_t),\,a_{\Att}(w_t),\,a_{\LM}(w_t),\,a(w_t)\,]^{\T},
\label{eq_aux}
\end{equation}
where
$a_{\ast}(w_t)$
is the score for token $w_t$,
which
can be obtained from the scores for partial token sequences
$w_{1:t}$ and $w_{1:t-1}$ as
$a_{\ast}(w_t)$ $=$
$a_{\ast}(w_{1:t})$ $-$ $a_{\ast}(w_{1:t-1})$.
%
\subsection{Class-balanced loss}
\label{ssec_cbloss}
%
As described above,
recent E2E ASR systems employ
state-of-the-art Transformers
and show high performance
for various ASR tasks
\cite{Karita_ASRU2019}.
In such situations,
training a confidence estimation model becomes difficult
since
the training data
contains
a large number of
correctly recognized (class 0) token
samples but
contains
a very small
number of
incorrectly recognized (class 1) token
samples.

In
model training
with
this
imbalanced dataset,
to avoid
underestimation (overestimation)
of the incorrect (correct) tokens,
we introduce class-balanced loss (CBLoss)
\cite{Cui_CVPR2019}.
Let
$\P^B$ $=$ $\{\p_1,\cdots,\p_B\}$
be the binary-class
posterior probability vector samples in a batch,
where
$\p_i$ $=$ $[p_{0,i},p_{1,i}]^{\T}$
(Eq.~(\ref{eq_pt}))
and $B$ is the number of samples in the batch.
$C^B$ $=$ $\{c_1,\cdots,c_B\}$
is
the
corresponding
binary-class samples (teacher labels).
%
Given $\P^B$ and $C^B$,
the class-balanced cross-entropy loss for the batch is obtained as,
\begin{equation}
\L^{\CB}(\P^B,C^B)
=-\sum_{i=1}^B
\frac{1-\beta}{1-\beta^{N_{c_i}}}
\log p_{c_i,i},
\label{eq_cbloss}
\end{equation}
where $\beta$
($0$ $\leq$ $\beta$ $<$ $1$)
is a tunable parameter
and $N_{c_i}$ is the total number of samples of class $c_i$ tokens
in the training data.
If $\beta$ is set
at
a large value (close to 1),
the class imbalance is more emphasized in the loss calculation,
i.e., a larger weight is applied to the incorrect (class 1) token samples
than the correct (class 0) token samples.
$\beta$ is usually set
in the range of $0.9\sim0.9999$
\cite{Cui_CVPR2019}.
In the implementation,
to make the class-balanced (weighted) loss
roughly
on
the same scale
with
the original (non-weighted) loss,
as proposed in \cite{Cui_CVPR2019},
we adjust the calculated binary-class weights
to satisfy their sum equals 2 (the number of classes).
%
\subsection{Transformer-based model}
\label{ssec_trans}
%
Transformers
\cite{Vaswani_NIPS2017}
have also been applied to sequence labeling problems,
such as part-of-speech (POS) tagging and named entity recognition (NER)
\cite{Guo_NAACLHLT2019,Yan_arXiv2019}.
In addition to the above BLSTM-based model,
in this study,
we also employ a Transformer-based
sequence labeler as a confidence estimation model.

In the BLSTM-based model
(Section~\ref{ssec_blstm}),
binary-class posterior probability vector $\p_t$
at time step $t$
is
estimated using forward and backward hidden state vectors
$\h_t^{\overrightarrow{}}$ and
$\h_t^{\overleftarrow{}}$,
in which the forward and backward contextual information
is
encoded.
In the Transformer-based model,
in contrast,
based on its self-attention operations,
the hidden state vectors of all the time steps,
$\h_1,\cdots,\h_T$,
are simultaneously utilized to estimate $\p_t$.
We
investigate which mechanism is suitable for confidence estimation.
%
\vspace{-1.50mm}
\section{Relation to Prior Work}
\label{sec_rel}
\vspace{-1.00mm}
%
As described in Section~\ref{sec_intro},
to the best of our knowledge,
only
a few studies
investigated
confidence estimation for E2E ASR systems
\cite{Woodward_IS2020,Kumar_IS2020}.
These are very informative studies.

The main differences between
them and ours
can be
summarized
as follows.
%
(B)LSTM-based E2E ASR systems are used
in \cite{Woodward_IS2020,Kumar_IS2020},
in contrast,
we use a
state-of-the-art
Transformer-based E2E ASR system
\cite{Karita_ASRU2019}
(Section~\ref{ssec_aux}).
Although
MLPs are used as confidence estimation models
in \cite{Woodward_IS2020,Kumar_IS2020},
we use stronger BLSTM-based models
(Section~\ref{ssec_blstm}).
Rich auxiliary features,
such as
ASR encoder output vectors
and ASR decoder state vectors,
which may be slightly difficult to extract,
are used
in \cite{Woodward_IS2020,Kumar_IS2020}.
In contrast,
we use easily available ASR decoding scores
as auxiliary features
(Section~\ref{ssec_aux}).
Since we use a Transformer-based ASR model
that shows high ASR performance,
we
need
to tackle a
severer
imbalanced
problem
than \cite{Woodward_IS2020,Kumar_IS2020}
and
we use
CBLoss
for stable model training
(Section~\ref{ssec_cbloss}).
%
\vspace{-1.25mm}
\section{Experiments}
\label{sec_exp}
\vspace{-1.25mm}
%
We conducted confidence estimation experiments
based on an ESPnet ASR recipe
using
the corpus of spontaneous Japanese (CSJ)
\cite{Maekawa_SSPR2003},
which is a large scale lecture speech corpus.
In the ESPnet CSJ ASR recipe
\cite{ESPnet},
a token corresponds to a character
(a Japanese character sequence includes both phonogram
and ideograph characters).
Such
details as the ASR model structure
and
the training/decoding settings
can be found in
the recipe
\cite{ESPnet}.
We used PyTorch
\cite{Paszke_NeurIPS2019}
for all the NN modeling in our study.
%
\begin{table}[t]
\vspace{-2.0mm}
\small
\caption{\blskip Details of training/validation/evaluation datasets.}
\label{tbl_data}
\vspace{0.5mm}
\begin{center}
\begin{tabular}{lrrr}
\hline
                                  &   Train &  Valid &   Eval \\ \hline
Hours                             &   249.7 &    6.4 &    5.1 \\
\# utterances (sentences)         &  199023 &   4000 &   3949 \\
\# characters                     & 5800994 & 151081 & 114795 \\
\# correct (class 0) characters   & 5503696 & 145531 & 110001 \\
\# incorrect (class 1) characters &  297298 &   5550 &   4794 \\
Incorrect character rate [\%]     &     5.1 &    3.7 &    4.2 \\
ASR character error rate [\%]     &     6.9 &    5.1 &    5.8 \\ \hline
\end{tabular}
\end{center}
\small
\caption{\blskip Structures of six BLSTM-based confidence estimation models.}
\label{tbl_blstm}
\vspace{-2.5mm}
\begin{center}
\begin{tabular}{ll}
\hline
Vocabulary size        & 3052 (\# characters)                           \\
Char embedding dims
& 8,\,\,\,\,16,\,\,\,\,32,\,\,\,\,64,\,\,\,\,128,\,\,\,\,256            \\
Auxiliary feature dims\,\, & 4 (Eq.~(\ref{eq_aux}))                     \\
\# LSTM hidden nodes   & Char embed dims + 4 (aux feat dims)            \\
\# LSTM layers         & 2                                              \\
\# linear hidden nodes & \# LSTM hidden nodes $\times$ 2 (fw\,+\,bw)    \\
\# linear layers       & 1                                              \\
Softmax size           & 2 (\# classes)                                 \\
\# parameters [M]
& 0.03,\,\,\,\,0.06,\,\,\,\,0.14,\,\,\,\,0.36,\,\,\,\,1.05,\,\,\,\,3.42 \\
\hline
\end{tabular}
\end{center}
\vspace{-3.516mm}
\end{table}
%
\vspace{-1.75mm}
\subsection{Experimental settings}
\label{ssec_set}
\vspace{-0.75mm}
%
As described in Section~\ref{ssec_aux},
we
employed
a Transformer-based encoder-decoder model
as the attention-based ASR model
\cite{Karita_ASRU2019}
along
with a CTC-based sub-decoder
\cite{Kim_ICASSP2017}.
We also used an LSTMLM with shallow fusion
fashion
\cite{Hori_IS2017}
as shown in Eq.~(\ref{eq_wsum}).
Although
we trained this ASR model based on the ESPnet
recipe
\cite{ESPnet},
we divided the original training data into two parts
to prepare the training data for confidence estimation models.
As a result,
the ASR model training data consists of
266 hours,
210k utterances (sentences),
and
6.14M characters.
Using this
half-sized
training data,
we trained the ASR model and the LSTMLM.

Using the ASR model and the LSTMLM,
we performed ASR for the training data of the confidence estimation models,
the validation data, and the evaluation data
shown in Table~\ref{tbl_data}
(the original recipe has
the three evaluation datasets,
but we merged them for simplicity)
and obtained ASR hypotheses (recognized character sequences)
with corresponding
four-dimensional
auxiliary feature vector sequences.
%
Decoding parameters $\lambda$ and $\gamma$ in Eq.~(\ref{eq_wsum})
were both set at 0.3.
Then we aligned the recognized character sequences
with the reference character sequences
and obtained the binary-class (correct/incorrect) symbol sequences
(teacher label sequences).
As shown in Table~\ref{tbl_data},
even though we halved the training data,
we obtained very low
ASR character rates (CERs)
(when we used the whole training data,
the CERs for the validation and evaluation datasets were 4.7\% and 5.2\%,
respectively).
As a result,
the incorrect character rates are
also
very low
and thus
these datasets are highly imbalanced
(note that the incorrectly recognized characters
correspond to
substitution or insertion
errors
%
and no deletion errors are included,
and thus
the incorrect character rates are always lower
than the ASR CERs, in which
deletion
errors
are counted).

We trained six BLSTM-based confidence estimation models
using the training data shown in Table~\ref{tbl_data}.
They have 
different model sizes
based on
their different character embedding
dimensions
shown in Table~\ref{tbl_blstm}.
%
We set
$\beta$ parameter of
CBLoss
in Eq.~(\ref{eq_cbloss}) at 0.9999
(four 9s)
based on
\cite{Cui_CVPR2019}.
With this $\beta$ value,
the weights for the correct (class 0)
samples and
the incorrect (class 1)
samples were set
at 0.99 and 1.01.
%
We used the Adam optimizer
\cite{Kingma_ICLR2015}
with a method for
scheduling the learning rate
\cite{Vaswani_NIPS2017}.
We set the batch size at 20 character sequences
and repeated the training for 20 epochs.
Among the 20 epoch models,
we selected
one
that shows the lowest loss for the validation data
in Table~\ref{tbl_data}
as the evaluation model.

With the same training procedure,
we trained an MLP
as the baseline model
\cite{Woodward_IS2020,Kumar_IS2020}.
%
It
consists of
a 32-dimensional character embedding layer,
six linear layers,
each
of which
has
36 nodes
(the embedding
plus
auxiliary feature vector dimensions)
with the ReLU activation function,
and a binary-class softmax layer.
This is our best configured MLP.
We also trained six Transformer-based models.
They
consist of
character embedding layers
whose
dimensions
are
shown in Table~\ref{tbl_blstm},
two layers,
two heads,
and a binary-class softmax layer.
Each layer consists of
the
embedding
dimensions
plus four
(the auxiliary feature vector dimensions)
nodes of a query, key, value, feed-forward,
and residual connection
networks.

Using the trained BLSTM and Transformer-based models,
we performed confidence estimation for the validation and evaluation datasets
shown in Table~\ref{tbl_data}.
For the performance
measurement,
we employed
three
commonly used
metrics,
i.e.,
the equal error rate (EER),
the area under the receiver operating characteristics curve (AUC),
and the normalized cross-entropy (NCE).
A good model shows a lower EER value
and higher AUC/NCE values.
The details of these metrics can be found in
\cite{Del-Agua_IEEEACMTASLP2018,Woodward_IS2020,Kumar_IS2020,Siu_CSL1999}.
%
\vspace{-2.50mm}
\subsection{Results of MLP and BLSTM-based models}
\label{ssec_res_blstm}
\vspace{-1.25mm}
%
The upper half of
Table~\ref{tbl_res} shows the
results
obtained with the MLP and the six BLSTM-based models.
We can confirm that
the BLSTM-based models
steadily show high confidence estimation performance
even under highly imbalanced conditions
and greatly outperform the MLP baseline.
Among the BLSTM-based models,
the smaller models outperform the larger models.
%
This
trend
can be attributable to
the small vocabulary size (3052 characters)
and the small softmax output size (two-classes)
as
shown in Table~\ref{tbl_blstm}.
%
\vspace{-2.50mm}
\subsection{Effects of auxiliary features and class-balanced loss}
\label{ssec_effects}
\vspace{-1.25mm}
%
Table~\ref{tbl_res_aux} shows the
results
obtained
with
the
BLSTM-based model,
which has the 16-dimensional character embedding layer,
without using the auxiliary features
(i.e., using only the character embeddings)
or using just one of them.
We can confirm that
the CTC score is
the most effective feature.
The CTC score is calculated
over an encoder output hidden state vector sequence
with
the forward-backward algorithm
\cite{Graves_ICML2014,Watanabe_IEEEACMTASLP2017}.
This
resembles
the confidence score
calculation over a word lattice
with
the forward-backward algorithm
\cite{Wessel_IEEETASLP2001}.
%
%
Therefore,
the CTC score itself can be understood as a kind of confidence score.
%
These four scores 
complement each other
and the
confidence estimation
performance
of the model
is greatly improved by using all of them simultaneously.

The second line from the bottom
in Table~\ref{tbl_res_aux}
(W/ all scores)
shows the results
obtained using
CBLoss by setting 
$\beta$ at 0.9999 (four 9s)
\cite{Cui_CVPR2019}
as described in Section~\ref{ssec_set}.
These results are almost the same
as
the results
obtained without using
CBLoss,
which is
where we
used
the
equal
weight (1.0)
for both the correct/incorrect classes.
The bottom line
in Table~\ref{tbl_res_aux}
shows
the results
when we
used
CBLoss by setting $\beta$ at 0.99999 (five 9s).
%
%
In this case,
we set the binary-class weights
at 0.97 and 1.03.
%
With
these results,
unfortunately,
we cannot confirm the effectiveness of introducing
CBLoss
(or we observe a slight performance degradation by setting a larger $\beta$).
We need to further investigate
the methods
for weighting
imbalanced samples.
%
\vspace{-2.50mm}
\subsection{Results of Transformer-based models}
\label{ssec_res_trans}
\vspace{-1.25mm}
%
The lower half of
Table~\ref{tbl_res} shows the
results
obtained with the six Transformer-based models.
We can confirm that
the performance of the Transformer-based models is inferior to
that of the BLSTM-based models.
Since
a similar trend is also reported in
\cite{Guo_NAACLHLT2019,Yan_arXiv2019}
for POS tagging and NER tasks,
the self-attention mechanism
over hidden state vectors of all the time steps
(Section~\ref{ssec_trans})
may be unsuitable for
sequence labeling problems.
The performance
gets
degraded along with the model size
increases,
and we fail to train the largest size model.
%
We analyzed the results
and
found that the models
greatly
underestimate (overestimate)
the incorrect (correct) samples.
We need to further investigate how to stably optimize 
the Transformer-based models.
%
\begin{table}[t]
\vspace{-2.0mm}
\small
\caption{\blskip Results obtained with MLP (baseline), six BLSTM-based models, and six Transformer-based models.}
\label{tbl_res}
\vspace{-2.5mm}
\begin{center}
\begin{tabular}{rrcccccc}
\hline
& Char
& \multicolumn{3}{c}{Validation}
& \multicolumn{3}{c}{Evaluation}\\
Model\!
& \hspace{-0.3mm}embed
&  \makebox[4.9mm]{EER} & \makebox[4.9mm]{AUC} & \makebox[4.9mm]{NCE}\,
&  \makebox[4.9mm]{EER} & \makebox[4.9mm]{AUC} & \makebox[4.9mm]{NCE}\\
\hline
\hspace{3.3mm} MLP\!
& 32
& \makebox[4.9mm]{12.7} & \makebox[4.9mm]{0.936} & \makebox[4.9mm]{0.395}\,
& \makebox[4.9mm]{14.3} & \makebox[4.9mm]{0.923} & \makebox[4.9mm]{0.364}\\
\hline
\hspace{3.3mm} BLSTM\!
& 8
& \makebox[4.9mm]{10.0}
& \makebox[4.9mm]{\bf 0.959}
& \makebox[4.9mm]{0.478}\,
& \makebox[4.9mm]{11.7}
& \makebox[4.9mm]{\bf 0.947}
& \makebox[4.9mm]{0.441}\\
& 16
& \makebox[4.9mm]{\bf\,\,\,9.9}
& \makebox[4.9mm]{\bf 0.959}
& \makebox[4.9mm]{\bf 0.483}\,
& \makebox[4.9mm]{\bf 11.6}
& \makebox[4.9mm]{\bf 0.947}
& \makebox[4.9mm]{\bf 0.442}\\
& 32
& \makebox[4.9mm]{\bf\,\,\,9.9}
& \makebox[4.9mm]{\bf 0.959}
& \makebox[4.9mm]{0.479}\,
& \makebox[4.9mm]{11.8}
& \makebox[4.9mm]{\bf 0.947}
& \makebox[4.9mm]{0.439}\\
& 64
& \makebox[4.9mm]{10.2} & \makebox[4.9mm]{0.957} & \makebox[4.9mm]{0.474}\,
& \makebox[4.9mm]{11.8} & \makebox[4.9mm]{0.946}
& \makebox[4.9mm]{0.438}\\
& 128
& \makebox[4.9mm]{10.3} & \makebox[4.9mm]{0.956} & \makebox[4.9mm]{0.473}\,
& \makebox[4.9mm]{12.1} & \makebox[4.9mm]{0.945} & \makebox[4.9mm]{0.435}\\
& 256
& \makebox[4.9mm]{10.4} & \makebox[4.9mm]{0.956} & \makebox[4.9mm]{0.470}\,
& \makebox[4.9mm]{12.4} & \makebox[4.9mm]{0.943} & \makebox[4.9mm]{0.428}\\
\hline
Transfomer\!
& 8
& \makebox[4.9mm]{12.5} & \makebox[4.9mm]{0.937} & \makebox[4.9mm]{0.401}\,
& \makebox[4.9mm]{14.0} & \makebox[4.9mm]{0.925} & \makebox[4.9mm]{0.367}\\
& 16
& \makebox[4.9mm]{12.4} & \makebox[4.9mm]{0.937} & \makebox[4.9mm]{0.402}\,
& \makebox[4.9mm]{13.9} & \makebox[4.9mm]{0.925} & \makebox[4.9mm]{0.370}\\
& 32
& \makebox[4.9mm]{12.9} & \makebox[4.9mm]{0.934} & \makebox[4.9mm]{0.399}\,
& \makebox[4.9mm]{14.4} & \makebox[4.9mm]{0.921} & \makebox[4.9mm]{0.363}\\
& 64
& \makebox[4.9mm]{13.7} & \makebox[4.9mm]{0.928} & \makebox[4.9mm]{0.384}\,
& \makebox[4.9mm]{15.7} & \makebox[4.9mm]{0.915} & \makebox[4.9mm]{0.350}\\
& 128
& \makebox[4.9mm]{14.2} & \makebox[4.9mm]{0.925} & \makebox[4.9mm]{0.377}\,
& \makebox[4.9mm]{16.1} & \makebox[4.9mm]{0.912} & \makebox[4.9mm]{0.344}\\
& 256
& \makebox[4.9mm]{17.0} & \makebox[4.9mm]{0.902} & \makebox[4.9mm]{0.333}\,
& \makebox[4.9mm]{18.5} & \makebox[4.9mm]{0.888} & \makebox[4.9mm]{0.303}\\
\hline
\end{tabular}
\end{center}
\vspace{-3.0mm}
\small
\caption{\blskip Results obtained with BLSTM-based model, which has a 16-dimensional character embedding layer, without using auxiliary features (i.e., using only character embeddings) or using just one auxiliary feature. The bottom line shows results obtained using CBLoss by setting $\beta$ at 0.99999 (five 9s).}
\label{tbl_res_aux}
\vspace{-2.5mm}
\begin{center}
\begin{tabular}{lcccccc}
\hline
& \multicolumn{3}{c}{Validation}
& \multicolumn{3}{c}{Evaluation}\\
Auxiliary features
& \makebox[4.9mm]{EER} & \makebox[4.9mm]{AUC} & \makebox[4.9mm]{NCE}\,
& \makebox[4.9mm]{EER} & \makebox[4.9mm]{AUC} & \makebox[4.9mm]{NCE}\\
\hline
Only char embedds
& \makebox[4.9mm]{29.4} & \makebox[4.9mm]{0.778} & \makebox[4.9mm]{0.117}\,
& \makebox[4.9mm]{30.2} & \makebox[4.9mm]{0.762} & \makebox[4.9mm]{0.103}\\
\hline
W/ CTC score
& \makebox[4.9mm]{\bf 12.2}
& \makebox[4.9mm]{\bf 0.941}
& \makebox[4.9mm]{\bf 0.405}\,
& \makebox[4.9mm]{\bf 13.6}
& \makebox[4.9mm]{\bf 0.928}
& \makebox[4.9mm]{\bf 0.371}\\
W/ attention score
& \makebox[4.9mm]{13.1} & \makebox[4.9mm]{0.936} & \makebox[4.9mm]{0.402}\,
& \makebox[4.9mm]{14.5} & \makebox[4.9mm]{0.922} & \makebox[4.9mm]{0.369}\\
W/ LSTMLM score
& \makebox[4.9mm]{18.6} & \makebox[4.9mm]{0.891} & \makebox[4.9mm]{0.279}\,
& \makebox[4.9mm]{21.5} & \makebox[4.9mm]{0.861} & \makebox[4.9mm]{0.227}\\
W/ weight sum score\!
& \makebox[4.9mm]{13.4} & \makebox[4.9mm]{0.936} & \makebox[4.9mm]{0.398}\,
& \makebox[4.9mm]{15.6} & \makebox[4.9mm]{0.919} & \makebox[4.9mm]{0.354}\\
\hline
W/ all scores
& \makebox[4.9mm]{\,\,\,9.9}
& \makebox[4.9mm]{0.959}
& \makebox[4.9mm]{0.483}\,
& \makebox[4.9mm]{11.6} & \makebox[4.9mm]{0.947} & \makebox[4.9mm]{0.442}\\
\hline
$\beta\!=\!0.99999$ (five 9s)\!
& \makebox[4.9mm]{\,\,\,9.9}
& \makebox[4.9mm]{0.959}
& \makebox[4.9mm]{0.480}\,
& \makebox[4.9mm]{11.8} & \makebox[4.9mm]{0.947} & \makebox[4.9mm]{0.440}\\
\hline
\end{tabular}
\end{center}
\vspace{-6.50mm}
\end{table}
%
\vspace{-0.75mm}
\section{Conclusion and Future Work}
\label{sec_con}
\vspace{-1.25mm}
%
We performed confidence estimation for an E2E ASR system
by using a BLSTM-based confidence estimation model.
Even under highly imbalanced experimental settings,
the BLSTM-model showed a high confidence estimation performance
by utilizing ASR decoding scores as auxiliary features.
We believe that
the experimental results obtained in this study
will be very informative in the research area of confidence estimation.

Future work will include
the use of richer features
\cite{Woodward_IS2020,Kumar_IS2020},
other class weighting methods
for the loss calculation,
e.g.,
focal loss
\cite{Lin_ICCV2017},
in combination with
CBLoss
\cite{Cui_CVPR2019},
pretraining methods
\cite{Devlin_NAACLHLT2019,Brown_arXiv2020}
for the stable training of the Transformer-based models,
and
conducting
experiments on
other ASR tasks
\cite{Karita_ASRU2019,ESPnet}.

%


%

\baselineskip 9.10pt
\bibliographystyle{IEEEtran}
\bibliography{ogawa}

\begin{thebibliography}{10}
\providecommand{\url}[1]{#1}
\csname url@samestyle\endcsname
\providecommand{\newblock}{\relax}
\providecommand{\bibinfo}[2]{#2}
\providecommand{\BIBentrySTDinterwordspacing}{\spaceskip=0pt\relax}
\providecommand{\BIBentryALTinterwordstretchfactor}{4}
\providecommand{\BIBentryALTinterwordspacing}{\spaceskip=\fontdimen2\font plus
\BIBentryALTinterwordstretchfactor\fontdimen3\font minus
  \fontdimen4\font\relax}
\providecommand{\BIBforeignlanguage}[2]{{%
\expandafter\ifx\csname l@#1\endcsname\relax
\typeout{** WARNING: IEEEtran.bst: No hyphenation pattern has been}%
\typeout{** loaded for the language `#1'. Using the pattern for}%
\typeout{** the default language instead.}%
\else
\language=\csname l@#1\endcsname
\fi
#2}}
\providecommand{\BIBdecl}{\relax}
\BIBdecl

\bibitem{Hinton_IEEESPM2012}
G.~Hinton, L.~Deng, D.~Yu, G.~E. Dahl, A.~rahman Mohamed, N.~Jaitly, A.~Senior,
  V.~Vanhoucke, P.~Nguyen, T.~N. Sainath, and B.~Kingsbury, ``Deep neural
  networks for acoustic modeling in speech recognition: {The} shared views of
  four research groups,'' \emph{IEEE Signal Processing Magazine}, vol.~29,
  no.~6, pp. 82--97, Nov. 2012.

\bibitem{Yu_Springer2015}
D.~Yu and L.~Deng, \emph{Automatic speech recognition: {A} deep learning
  approach}.\hskip 1em plus 0.5em minus 0.4em\relax Springer-Verlag, London,
  2015.

\bibitem{Jiang_SPECOM2005}
H.~Jiang, ``Confidence measures for speech recognition: {A} survey,''
  \emph{Speech Communication}, vol.~45, no.~4, pp. 455--470, April 2005.

\bibitem{San-Segundo_ICASSP2001}
R.~San-Segundo, B.~Pellom, K.~Hacioglu, and W.~Ward, ``Confidence measures for
  spoken dialogue systems,'' in \emph{Proc. ICASSP}, 2001, pp. 393--396.

\bibitem{Swarup_IS2019}
P.~Swarup, R.~Maas, S.~Garimella, S.~H. Mallidi, and B.~Hoffmeister,
  ``Improving {ASR} confidence scores for {Alexa} using acoustic and hypothesis
  embeddings,'' in \emph{Proc. Interspeech}, 2019, pp. 2175--2179.

\bibitem{Lleida_IEEETSAP2000}
E.~Lleida and R.~C. Rose, ``Utterance verification in continuous speech
  recognition: {Decoding} and training procedures,'' \emph{IEEE Transactions on
  Speech and Audio Processing}, vol.~8, no.~2, pp. 126--139, 2000.

\bibitem{Wessel_IEEETASLP2001}
F.~Wessel, R.~Schl{\"u}ter, K.~Macherey, and H.~Ney, ``Confidence measures for
  large vocabulary continuous speech recognition,'' \emph{IEEE Transactions on
  Audio, Speech, and Language Processing}, vol.~9, no.~3, pp. 288--298, 2001.

\bibitem{Mangu_CSL2000}
L.~Mangu, E.~Brill, and A.~Stolcke, ``Finding consensus in speech recognition:
  {Word} error minimization and other applications of confusion networks,''
  \emph{Computer Speech and Language}, vol.~14, no.~4, pp. 373--400, Oct. 2000.

\bibitem{Fayolle_IS2010}
J.~Fayolle, F.~Moreau, C.~Raymond, G.~Gravier, and P.~Gros, ``{CRF}-based
  combination of contextual features to improve a posteriori word-level
  confidence measures,'' in \emph{Proc. Interspeech}, 2010, pp. 1942--1945.

\bibitem{Seigel_IS2011}
M.~S. Seigel and P.~C. Woodland, ``Combining information sources for confidence
  estimation with {CRF} models,'' in \emph{Proc. Interspeech}, 2011, pp.
  905--908.

\bibitem{Yu_IEEETASLP2011}
D.~Yu, J.~Li, and L.~Deng, ``Calibration of confidence measures in speech
  recognition,'' \emph{IEEE Transactions on Audio, Speech, and Language
  Processing}, vol.~19, no.~8, pp. 2461--2473, Nov. 2011.

\bibitem{Schaaf_ICASSP1997}
T.~Schaaf and T.~Kemp, ``Confidence measures for spontaneous speech
  recognition,'' in \emph{Proc. ICASSP}, 1997, pp. 875--878.

\bibitem{Weintraub_ICASSP1997}
M.~Weintraub, F.~Beaufays, Z.~Rivlin, Y.~Konig, and A.~Stolcke,
  ``Neural-network based measures of confidence for word recognition,'' in
  \emph{Proc. ICASSP}, 1997, pp. 887--890.

\bibitem{Tam_ICASSP2014}
Y.-C. Tam, Y.~Lei, J.~Zheng, and W.~Wang, ``{ASR} error detection using
  recurrent neural network language model and complementary {ASR},'' in
  \emph{Proc. ICASSP}, 2014, pp. 2331--2335.

\bibitem{Kalgaonkar_ICASSP2015}
K.~Kalgaonkar, C.~Liu, Y.~Gong, and K.~Yao, ``Estimating confidence scores on
  {ASR} results using recurrent neural networks,'' in \emph{Proc. ICASSP},
  2015, pp. 4999--5003.

\bibitem{Ogawa_ICASSP2015}
A.~Ogawa and T.~Hori, ``{ASR} error detection and recognition rate estimation
  using deep bidirectional recurrent neural networks,'' in \emph{Proc. ICASSP},
  2015, pp. 4370--4374.

\bibitem{Ogawa_SPECOM2017}
------, ``Error detection and accuracy estimation in automatic speech
  recognition using deep bidirectional recurrent neural networks,''
  \emph{Speech Communication}, vol.~89, pp. 70--83, May 2017.

\bibitem{Del-Agua_IEEEACMTASLP2018}
M.~{\'A}. Del-Agu, A.~Gim{\'e}nez, A.~Sanchis, J.~Civera, and A.~Juan,
  ``Speaker-adapted confidence measures for {ASR} using deep bidirectional
  recurrent neural networks,'' \emph{IEEE/ACM Transactions on Audio, Speech,
  and Language Processing}, vol.~26, no.~7, pp. 1198--1206, 2018.

\bibitem{Hochreiter_NeuralComput1997}
S.~Hochreiter and J.~Schmidhuber, ``Long short-term memory,'' \emph{Neural
  Computation}, vol.~9, no.~8, pp. 1735--1780, Nov. 1997.

\bibitem{Schuster_IEEETSP1997}
M.~Schuster and K.~K. Paliwal, ``Bidirectional recurrent neural networks,''
  \emph{IEEE Transactions on Signal Processing}, vol.~45, no.~11, pp.
  2673--2681, Nov. 1997.

\bibitem{Chorowski_arXiv2014}
J.~Chorowski, D.~Bahdanau, K.~Cho, and Y.~Bengio, ``End-to-end continuous
  speech recognition using attention-based recurrent {NN}: {First} results,''
  \emph{arXiv:1412.1602 [cs.NE]}, 2014.

\bibitem{Chan_ICASSP2016}
W.~Chan, N.~Jaitly, Q.~Le, and O.~Vinyals, ``Listen, attend and spell: {A}
  neural network for large vocabulary conversational speech recognition,'' in
  \emph{Proc. ICASSP}, 2016, pp. 4960--4964.

\bibitem{Watanabe_IS2018}
S.~Watanabe, T.~Hori, S.~Karita, T.~Hayashi, J.~Nishitoba, Y.~Unno, N.~E.~Y.
  Soplin, J.~Heymann, M.~Wiesner, N.~Chen, A.~Renduchintala, and T.~Ochiai,
  ``{ESP}net: {End}-to-end speech processing toolkit,'' in \emph{Proc.
  Interspeech}, 2018, pp. 2207--2211.

\bibitem{Ott_NAACLHLT2019}
M.~Ott, S.~Edunov, A.~Baevski, A.~Fan, S.~Gross, N.~Ng, D.~Grangier, and
  M.~Auli, ``{FAIRSEQ}: {A} fast, extensible toolkit for sequence modeling,''
  in \emph{Proc. NAACL-HLT: Demonstrations}, 2019, pp. 48--53.

\bibitem{Karita_ASRU2019}
S.~Karita, N.~Chen, T.~Hayashi, T.~Hori, H.~Inaguma, Z.~Jiang, M.~Someki,
  N.~E.~Y. Soplin, R.~Yamamoto, X.~Wang, S.~Watanabe, T.~Yoshimura, and
  W.~Zhang, ``A comparative study on {Transformer} vs {RNN} in speech
  applications,'' in \emph{Proc. ASRU}, 2019, pp. 449--456.

\bibitem{Woodward_IS2020}
A.~Woodward, C.~Bonn{\'i}n, I.~Masuda, D.~Varas, E.~Bou-Balust, and J.~C.
  Riveiro, ``Confidence measures in encoder-decoder models for speech
  recognition,'' in \emph{Proc. Interspeech}, 2020.

\bibitem{Kumar_IS2020}
A.~Kumar, S.~Singh, D.~Gowda, A.~Garg, S.~Singh, and C.~Kim, ``Utterance
  confidence measure for end-to-end speech recognition with applications to
  distributed speech recognition scenarios,'' in \emph{Proc. Interspeech},
  2020.

\bibitem{Cui_CVPR2019}
Y.~Cui, M.~Jia, T.-Y. Lin, Y.~Song, and S.~Belongie, ``Class-balanced loss
  based on effective number of samples,'' in \emph{Proc. CVPR}, 2019, pp.
  9260--9269.

\bibitem{Vaswani_NIPS2017}
A.~Vaswani, N.~Shazeer, N.~Parmar, J.~Uszkoreit, L.~Jones, A.~N. Gomez, and
  {\L}.~Kaiser, ``Attention is all you need,'' in \emph{Proc. NIPS}, 2017, pp.
  5998--6008.

\bibitem{Guo_NAACLHLT2019}
Q.~Guo, X.~Qiu, P.~Liu, Y.~Shao, X.~Xue, and Z.~Zhang, ``Star-{Transformer},''
  in \emph{Proc. NAACL-HLT}, 2019, pp. 1315--1325.

\bibitem{Yan_arXiv2019}
H.~Yan, B.~Deng, X.~Li, and X.~Qiu, ``{TENER}: {Adapting} {Transformer} encoder
  for named entity recognition,'' \emph{arXiv:1911.04474 [cs.CL]}, 2019.

\bibitem{Graves_ICML2014}
A.~Graves and N.~Jaitly, ``Towards end-to-end speech recognition with recurrent
  neural networks,'' in \emph{Proc. ICML}, 2014, pp. 1764--1772.

\bibitem{Kim_ICASSP2017}
S.~Kim, T.~Hori, and S.~Watanabe, ``Joint {CTC}-attention based end-to-end
  speech recognition using multi-task learning,'' in \emph{Proc. ICASSP}, 2017,
  pp. 4835--4839.

\bibitem{Watanabe_IEEEACMTASLP2017}
S.~Watanabe, T.~Hori, S.~Kim, J.~R. Hershey, and T.~Hayashi, ``Hybrid
  {CTC}/{Attention} architecture for end-to-end speech recognition,''
  \emph{IEEE/ACM Transactions on Audio, Speech, and Language Processing},
  vol.~11, no.~8, pp. 1240--1253, Dec. 2017.

\bibitem{Hori_IS2017}
T.~Hori, S.~Watanabe, Y.~Zhang, and W.~Chan, ``Advances in joint
  {CTC}-attention based end-to-end speech recognition with a deep {CNN} encoder
  and {RNN-LM},'' in \emph{Proc. Interspeech}, 2017, pp. 949--953.

\bibitem{Maekawa_SSPR2003}
K.~Maekawa, ``Corpus of spontaneous {J}apanese: {I}ts design and evaluation,''
  in \emph{Proc. Workshop on Spontaneous Speech Processing and Recognition
  (SSPR)}, 2003, pp. 7--12.

\bibitem{ESPnet}
S.~Watanabe, ``{ESP}net: {End}-to-end speech processing toolkit,''
  https:{\slash}{\slash}github.com{\slash}espnet{\slash}espnet.

\bibitem{Paszke_NeurIPS2019}
A.~Paszke \emph{et~al.}, ``{P}y{T}orch: {A}n imperative style, high-performance
  deep learning library,'' in \emph{Proc. NeurIPS}, 2019, pp. 8024--8035.

\bibitem{Kingma_ICLR2015}
D.~P. Kingma and J.~L. Ba, ``Adam: {A} method for stochastic optimization,'' in
  \emph{Proc. ICLR}, 2015.

\bibitem{Siu_CSL1999}
M.~Siu and H.~Gish, ``Evaluation of word confidence for speech recognition
  systems,'' \emph{Computer Speech and Language}, vol.~13, no.~4, pp. 299--319,
  Oct. 1999.

\bibitem{Lin_ICCV2017}
T.-Y. Lin, P.~Goyal, R.~Girshick, K.~He, and P.~Doll{\'a}r, ``Focal loss for
  dense object detection,'' in \emph{Proc. ICCV}, 2017, pp. 2999--3007.

\bibitem{Devlin_NAACLHLT2019}
J.~Devlin, M.-W. Chang, K.~Lee, and K.~Toutanova, ``{BERT}: {Pre-training} of
  deep bidirectional {Transformers} for language understanding,'' in
  \emph{Proc. NAACL-HLT}, 2019, pp. 4171--4186.

\bibitem{Brown_arXiv2020}
T.~B. Brown \emph{et~al.}, ``Language models are few-shot learners,''
  \emph{arXiv:2005.14165 [cs.CL]}, 2020.

\end{thebibliography}

\end{document}